\documentstyle[aps,epsfig]{revtex}
\newcommand{\gsim}{\mathrel{\rlap{\lower4pt\hbox{\hskip0pt$\sim$}}
\raise1pt\hbox{$>$}}}
\newcommand{\lsim}{\mathrel{\rlap{\lower4pt\hbox{\hskip0pt$\sim$}}
\raise1pt\hbox{$<$}}}
\begin{document}
\twocolumn
\preprint{Preprint Number: \parbox[t]{45mm}{ANL-PHY-9413-TH-99\\
}}
 \title{$K \to \pi\pi$ and a light scalar meson}
\author{J. C. R. Bloch,\footnotemark[1] 
        M. A. Ivanov,\footnotemark[2] 
        T. Mizutani,\footnotemark[3] 
        C. D. Roberts\footnotemark[1] and
        S. M. Schmidt\footnotemark[1]\vspace*{0.2\baselineskip}}

 \address{
 \footnotemark[1]Physics Division, Argonne National Laboratory,
 Argonne IL 60439-4843\vspace*{0.2\baselineskip}\\
\footnotemark[2]Bogoliubov Laboratory of Theoretical Physics,
 \\Joint Institute for Nuclear Research, 141980 Dubna,
 Russia\vspace*{0.2\baselineskip} \\
\footnotemark[3]Department
 of Physics, Virginia Polytechnic Institute and State University,
 Blacksburg, VA 24061-0435\vspace*{0.2\baselineskip}}
\date{ANL-PHY-9413-TH-99; Pacs Numbers: 13.20.Eb, 13.30.Eg, 12.15.Hh, 24.85.+p}
\maketitle
\begin{abstract}
We explore the $\Delta I=\case{1}{2}$ rule and $\epsilon^\prime/\epsilon$ in
$K\to \pi\pi$ transitions using a Dyson-Schwinger equation model.  Exploiting
the feature that QCD penguin operators direct $K^0_S$ transitions through
$0^{++}$ intermediate states, we find an explanation of the enhancement of
$K\to \pi\pi_{I=0}$ transitions in the contribution of a light
$\sigma$-meson.  This mechanism also affects $\epsilon^\prime/\epsilon$.  
\end{abstract}
\vspace*{\baselineskip}
\section{Introduction}
The $\Delta I = \frac{1}{2}$ rule is an empirical observation: the widths for
nonleptonic decays of kaons and hyperons that change isospin by one-half unit
are significantly larger than those for other $K$ and $\Lambda$ transitions;
e.g.,\cite{pdg98}
\begin{equation}
\label{expta}
\Gamma_{K_S^0 \to (\pi\pi)}/\Gamma_{K^+\to \pi^+\pi^0} = 660\,.
\end{equation}
In terms of the amplitudes $M_{K^0_S\to\pi^+\pi^-}$ and
$M_{K^0_S\to\pi^0\pi^0}$ that describe $K_S^0 \to \pi\pi$ transitions, the
pure isospin-zero and isospin-two $\pi\,\pi$ final states are described by
\begin{eqnarray}
\label{A0}
A_0 & = & 
\case{1}{\sqrt{6}} ( 2 M_{K^0_S\to\pi^+\pi^-} + M_{K^0_S\to\pi^0\pi^0})\,,\\
\label{A2}
A_2 & = & 
\case{1}{\sqrt{3}} ( M_{K^0_S\to\pi^+\pi^-} - M_{K^0_S\to\pi^0\pi^0})\,,
\end{eqnarray}
and the ratio in Eq.~(\ref{expta}) corresponds to:
\begin{equation}
\label{wdef}
1/w := {\rm Re}(A_0)/{\rm Re}(A_2) \approx 22\,.
\end{equation}
The analogous amplitude ratio for $S$-wave $\Lambda \to \pi N$ transitions is
$|A_{1/2}/A_{3/2}| \approx 80$.

The processes involved are nonleptonic weak decays so one necessarily
encounters QCD effects in their analysis and the operator product expansion
(OPE) can therefore be employed to good effect.  Using the OPE the amplitude
for a given transition is expressed as the expectation value of an effective
Hamiltonian:
$ A = \langle {\cal H}_{\rm eff}\rangle = \sum_i\,a_i(\mu)\,\langle
Q_i(\mu)\rangle\,, $
where $\mu$ is a renormalisation point.  The coefficients: $a_i(\mu)$, are
calculable in perturbation theory and describe short-distance effects.
However, the expectation values of the local effective operators: $\langle
Q_i(\mu)\rangle$, contain the effects of bound state structure; i.e.,
long-distance QCD effects, and must be calculated nonperturbatively.

The transitions of interest herein are mediated by nonleptonic strangeness
changing ($\Delta S = 1$) effective operators.  The simplest:
\begin{eqnarray}
Q_1 & = & \bar s_i O_\mu^- u_j \; \bar u_j O_\mu^- d_i\,, \\
Q_2 & = & \bar s_i \,O_\mu^- u_i \; \bar u_j \,O_\mu^- d_j\,,
\end{eqnarray}
with $O_\mu^\pm = \gamma_\mu(1\pm\gamma_5)$ and colour indices:
$i,j=1,\ldots,N_c$, have the flavour structure of the standard weak
four-fermion current-current interaction, and there are eight other terms
representing the QCD and electroweak (ew) penguin operators.\footnote{$Q_1$
results from QCD corrections to the weak current-current vertex.  The penguin
operators are generated by QCD and ew corrections but their flavour structure
is different.}
At least some of these must have large expectation values if the $\Delta I =
\frac{1}{2}$ rule is to be understood.

Another quantity that may be much influenced by the $\Delta S=1$ effective
interaction is the ratio $\epsilon^\prime/\epsilon$.  The indirect CP
violating parameter:
\begin{equation}
\epsilon := A(K_L \to \pi\pi_{I=0})/A(K_S \to \pi\pi_{I=0})
\end{equation}
measures the admixture of CP-even state in $K_L$: for $\epsilon = 0$, ${\rm
CP}\,|K_{L/S}\rangle = \mp\, |K_{L/S}\rangle$; i.e., they are CP eigenstates.
$\epsilon$ appears to be primarily determined by short-distance contributions
from the weak nonleptonic $\Delta S=2$ effective interaction\cite{buras}.  In
contrast, $\epsilon^\prime$ measures the phase of the heavy-quark CKM matrix
elements in the standard model and
\begin{equation}
\label{ratio}
\frac{\epsilon^\prime}{\epsilon} = \frac{1}{\sqrt{2}\,
|\epsilon|}\, {\rm Im}\left(\frac{A_2}{A_0}\right)\,,
\end{equation}
with $|\epsilon| = 0.002280$, experimentally.\footnote{In Eq.~(\ref{ratio})
we follow contemporary practice and make explicit the $\pi\pi$-scattering
phase shifts: $\delta_{0,2}$, in factoring out the phase
$\Phi_{\epsilon^\prime} = \case{\pi}{2} + \delta_2 - \delta_0$.  Then, using
the experimental observations: $\delta_0\approx 37^\circ$, $\delta_2\approx
-7^\circ$, $\Phi_\epsilon \approx \case{\pi}{4}$, one has
$\Phi_{\epsilon^\prime}-\Phi_\epsilon \approx 0$ and the imaginary part in
Eq.~(\ref{ratio}) relates only to an explicit CP violating phase.}
A nonzero value of $\epsilon^\prime/\epsilon$ entails direct transitions
between CP-even and CP-odd eigenstates.  $\epsilon^\prime$ is sensitive to
the same penguin operators that contribute to the $\Delta I = \frac{1}{2}$
rule, and hence is likely to receive significant long-distance contributions.
The current generation of experiments\cite{expts} appears to be consistent
and an average value of the ratio is\cite{epe}
\begin{equation}
\label{expt}
{\rm Re}\left(\frac{\epsilon^\prime}{\epsilon}\right) 
= (2.1 \pm 0.46) \times 10^{-3}\,.
\end{equation}

A standard form of the $\Delta S=1$ effective interaction at a
renormalisation scale $\mu = 1\,$GeV is
\begin{equation}
\label{deltas}
{\cal H}^{\Delta S=1}_{\rm eff} = \tilde G_F \sum_{i=1}^{10}\,
c_i(\mu)\,Q_i(\mu)\,, 
\end{equation}
where 
$\tilde G_F = G_F\,V_{us}^\ast\,V_{ud}/\sqrt{2}$,
$c_i(\mu)= z_i(\mu) + \tau \, y_i(\mu)$, $\tau= - (V^\ast_{ts}
V_{td})/(V_{us}^\ast\,V_{ud})$, and $V_{ud}$,\ldots, are the CKM matrix
elements.  (Direct CP violation is a measure of Im$(\tau)$.)  The
coefficients: $c_i(\mu)$, at next-to-leading order are quoted in
Ref.~\cite{buras}, as are the operators: $Q_i$.  We reproduce the
coefficients in the appendix, Eq.~(\ref{coeffs}), but not the operators and
note only that $Q_{3,4,5,6}$ are the QCD penguin operators; e.g.,
\begin{equation}
Q_6 = \bar s_i O_\mu^- d_j
\sum_{q=u,d,s} \,\bar q_j O_\mu^+ q_i\,,
\end{equation}
and $Q_{7,8,9,10}$ are the
ew penguin operators; e.g.,
\begin{equation}
Q_{8} = \case{3}{2} \,\bar s_i O_\mu^- d_j
\sum_{q=u,d,s} e_q\,\bar q_j O_\mu^+ q_i\,,
\end{equation}
where $e_q$ is the quark's electric charge (in units of the positron charge).
The expectation value of the operators in Eq.~(\ref{deltas}); i.e., the
long-distance contributions, are the primary source of theoretical
uncertainty in the estimation of $w$ and $\epsilon^\prime/\epsilon$,
Eqs.~(\ref{wdef}) and~(\ref{ratio}).

Herein we calculate the expectation values of the operators in
Eq.~(\ref{deltas}) using the Dyson-Schwinger equation (DSE) model of
Ref.~\cite{misha}.  The DSE framework is reviewed in Ref.~\cite{cdragw} with
some of the phenomenological applications described in
Refs.~\cite{peterrev,echaya,nucleon,mike2,mt99}.  It treats mesons as bound
states of a dressed-quark and \mbox{-antiquark} with Bethe-Salpeter
amplitudes describing their internal structure, and has already been used to
explore CP violation in hadrons\cite{martin}.  We describe the calculation
and its elements in Sect.~\ref{sec2}, and present and discuss our results in
Sect.~\ref{sec3}.  Section~\ref{epil} is a brief recapitulation.

\section{Operator Expectation Values}
\label{sec2}
\subsection{Charged kaon decay}
The impulse approximation to the meson-meson transitions mediated by ${\cal
H}^{\Delta S=1}_{\rm eff}$ is straightforward to evaluate; e.g., in the
absence of ew penguins only the operators $Q_{1,2}$ contribute to
$K^+\to\pi^+\pi^0$ transitions and
\begin{eqnarray}
\langle \pi^+(p_1) \pi^0(p_2)|Q_1|K^+(p)\rangle & = &
\case{1}{\sqrt{2}} \sum_{i=1,2}N_c^i\,T_i(p_1,p_2)\,,
\end{eqnarray}\vspace*{-1.5\baselineskip}
\begin{eqnarray}
\lefteqn{T_1(p_1,p_2) = i\sqrt{2}\,{\rm tr}Z_2\!\int_{k_2}^\Lambda\!\!
O_\mu^-\,\chi_\pi(k_2;-\case{1}{2}p_2,-\case{1}{2}p_2) }\\ 
&\times& \nonumber 2\,{\rm tr}Z_2\!\int_{k_1}^\Lambda \!\!
O_\mu^-\chi_K(k_1;p_2,p_1)\,
\Gamma_\pi(k_1;-p_1)\,S_u(k_1)\,,\\
\nonumber \rule{0mm}{\parindent}\lefteqn{i\,T_2(p_1,p_2) =
2\sqrt{2}{\rm tr}Z_2^2\!\int_{k_1}^\Lambda \int_{k_2}^\Lambda\!\!
O_\mu^-\,\chi_\pi(k_2;-\case{1}{2}p_2,-\case{1}{2}p_2) }\\
&& 
 \times \,O_\mu^-\,\chi_K(k_1;p_2,p_1)\,\Gamma_\pi(k_1;-p_1)\,S_u(k_1)\,,
\end{eqnarray}
with the trace over Dirac indices only, and
\begin{eqnarray}
\label{chipi}
\chi_\pi(k;\ell_1,\ell_2)&=& S_u(k+ \ell_1)\, \Gamma_\pi(k;\ell_1+\ell_2)\,
S_u(k-\ell_2)\,,\\
\chi_K(k;\ell_1,\ell_2) &= &S_s(k+ \ell_1)\, \Gamma_K(k;\ell_1+\ell_2)\,
S_u(k-\ell_2)\,.
\end{eqnarray}
Here we use a Euclidean formulation with
$\{\gamma_\mu,\gamma_\nu\}=2\delta_{\mu\nu}$, $\gamma_\mu^\dagger =
\gamma_\mu$ and $p\cdot q=\sum_{i=1}^4 p_i q_i$.  $\int_{k}^\Lambda:=
\int^\Lambda d^4 k/(2\pi)^4$ is a mnemonic representing a translationally
invariant regularisation of the integral, with $\Lambda$ the regularisation
mass-scale that is removed ($\Lambda \to \infty$) as the final stage of any
calculation, and $Z_2(\mu,\Lambda)$ is the quark wave function
renormalisation constant.  $S_{f=u,s}$ are the dressed-quark propagators (we
assume isospin symmetry) and $\Gamma_{H=K,\pi}$ are the meson Bethe-Salpeter
amplitudes, both of which we discuss in detail in Sect.~\ref{propsec}.

Using the Fierz rearrangement property: 
\begin{equation}
{\rm tr}\!\left[O_\mu^- G_1 O_\mu^- G_2\right] = - {\rm tr}\left[O_\mu^-
G_1\right]\,{\rm tr}\!\left[ O_\mu^- G_2\right] \,,
\end{equation}
where $G_{1,2}$ are any Dirac matrices, it is clear that $T_1 \propto T_2$.
Furthermore, the analysis for $Q_2$ is similar and the result is the same so
that
\begin{eqnarray}
\nonumber
\lefteqn{\langle \pi^+(p_1) \pi^0(p_2)|(c_1 Q_1+ c_2 Q_2)|K^+(p)\rangle }\\
&= &
\frac{c_1+c_2}{\sqrt{2}}\, N_c (N_c+1)\,
T_1(p_1,p_2)\,.
\end{eqnarray}
This can be simplified using\cite{mr97}
\begin{eqnarray}
f_\pi p_\mu  & = & 
 -\sqrt{2} \,N_c {\rm tr}Z_2\!\int_{k}^\Lambda 
O_\mu^-\,\chi_\pi(k;-\case{1}{2}p,-\case{1}{2}p)
\end{eqnarray}
and\cite{kl3}
\begin{eqnarray}
\nonumber
\lefteqn{-(p+p_1)_\mu \,f_+^{K^+}(p_2^2) - p_{2\mu} \,f_-^{K^+}(p_2^2)
=  }\\
&& 2 N_c {\rm tr}Z_2\!\int_{k_1}^\Lambda\,i O_\mu^-
\chi_K(k_1;p_2,p_1)\,\Gamma_\pi(k_1;-p_1)\,S_u(k_1)\,,
\end{eqnarray}
where $f_\pm^{K^+}$ are the $K_{\ell 3}$ semileptonic transition form
factors, so that 
\begin{eqnarray}
\nonumber 
\lefteqn{\langle \pi^+(p_1) \pi^0(p_2)|{\cal H}_{\rm eff}^{\Delta
S=1}|K^+(p)\rangle  }\\ 
\label{melement} &= &  \frac{N_c + 1}{\sqrt{2} \,N_c} 
\,\tilde G_F \, (c_1 + c_2)\, {\cal M}_1(p_1,p_2)\,,\\
\nonumber \lefteqn{{\cal M}_1(p_1,p_2) = }\\
& & 
 f_\pi \left[p_2.(p+p_1) f_+^{K^+}(p_2^2) + p_2^2\, f_-^{K^-}(p_2^2)\right]\\
\label{vires} & \approx & f_\pi\,(m_K^2 - m_\pi^2)\,,
\end{eqnarray}
where the last line follows from\cite{kl3} $f_+^{K^+}(-m_\pi^2)\approx - 1.0$
and $m_\pi^2f_-^{K^-}(-m_\pi^2)\approx 0$.  

We can compare our result with the contemporary phenomenological approach to
$K\to\pi\pi$ decays, which employs a parametrisation of ${\cal M}_1$:
\begin{equation}
{\cal M}_1 = f_\pi\,(m_K^2 - m_\pi^2)\,B_1^{(3/2)} \,,
\end{equation}
with the parameter $B_1^{(3/2)}$ fixed by fitting the experimental width.
One historical means of estimating ${\cal M}_1$ is to employ the vacuum
saturation {\it Ansatz}, which yields $B_1^{(3/2)}=1$.  It is thus clear from
Eqs.~(\protect\ref{melement}) and (\protect\ref{vires}) that our impulse
approximation is equivalent to this {\it Ansatz}.\footnote{This is an exact
algebraic constraint, which has been overlooked by other authors and
consequently violated in fitting $\Gamma_{K^+ \to \pi^+ \pi^0}$; e.g.,
Ref.~\protect\cite{mishaold} and references therein.}  However, agreement
with the experimental value of $\Gamma_{K^+ \to \pi^+ \pi^0}$ requires
$B_1^{(3/2)}\approx \case{1}{2}$, as can be seen using Eq.~(\ref{kpppp0}).
Thus, while the impulse approximation is reliable for estimating the
order-of-magnitude, it appears that an accurate result requires additional
contributions.\footnote{The impulse approximation, with dressed-propagators
and vertices, has proven quantitatively reliable in many other applications;
e.g., Refs.~\protect\cite{misha,peterrev,echaya,nucleon,mike2,mt99}, and in
some cases corrections have been calculated and shown to be
small\cite{kl3,bender}.  The new feature herein is that the calculation is
not self-contained; i.e., we presently rely on external input: the $c_i$ in
Eq.~(\protect\ref{deltas}).}
However, our primary goal is to identify a plausible mechanism for an
enhancement of $\pi\pi_{I=0}$ transitions and this level of accuracy is
sufficient for that purpose.  Hence we proceed by adopting the contemporary
artifice and use
\begin{equation}
\label{b1fix}
{\cal M}_1(p_1,p_2) := f_\pi\,(m_K^2 -
m_\pi^2)\,B_1^{(3/2)},\;B_1^{(3/2)}=\case{1}{2}\,. 
\end{equation}
In doing this we bypass the calculation of $B_1^{(3/2)}$, which our elucidation
of the impulse approximation has identified as a real challenge for models
whose basis is kindred to ours, and also for other approaches.

\subsection{Propagators and Bethe-Salpeter amplitudes}
\label{propsec}
Although the matrix element discussed above was expressed in terms of dressed
$u$- and $s$-quark propagators, and $\pi$- and $K$-meson Bethe-Salpeter
amplitudes, we obtained a model independent result without introducing
specific forms.  That is an helpful but uncommon simplification only
encountered before in the study of anomalous processes such as $\pi^0\to
\gamma\gamma$\cite{mrpion}. 

In general, as reviewed in Ref.~\cite{cdragw}, these quantities can be
obtained as solutions of the quark DSE and meson Bethe-Salpeter equation.
However, the successful study of an extensive range of low- and high-energy
light- and heavy-quark
phenomena\cite{misha,peterrev,echaya,nucleon,mike2,mt99,martin} has led to
the development of efficacious algebraic parametrisations.  These were
employed in Ref.~\cite{misha} and we also use them herein.

The dressed-quark propagator is
\begin{eqnarray}
\label{qprop}
S_f(p) & = & -i\gamma\cdot p\, \sigma_V^f(p^2) + \sigma_S^f(p^2)\,,\\
& = & \label{defS}
\left[i \gamma\cdot p \, A_f(p^2) + B_f(p^2)\right]^{-1}\,,\\
\label{ssm}
\bar\sigma_S^f(x) & =&  2\,\bar m_f \,{\cal F}(2 (x+\bar m_f^2))\\
&& \nonumber
+ {\cal F}(b_1^f x) \,{\cal F}(b_3^f x) \,
\left[b_0^f + b_2^f {\cal F}(\varepsilon x)\right]\,,\\
\label{svm}
\bar\sigma_V^f(x) & = & \frac{1}{x+\bar m_f^2}\,
\left[ 1 - {\cal F}(2 (x+\bar m_f^2))\right]\,,
\end{eqnarray}
with ${\cal F}(y) = (1-{\rm e}^{-y})/y$, $x=p^2/\lambda^2$, $\bar m_f$ =
$m_f/\lambda$, $\bar\sigma_S^f(x) = \lambda\,\sigma_S^f(p^2)$ 
$\bar\sigma_V^f(x) = \lambda^2\,\sigma_V^f(p^2)$.  The mass-scale,
$\lambda=0.566\,$GeV, and parameter values:\footnote{$\varepsilon=10^{-4}$ in
Eq.~(\ref{ssm}) acts only to decouple the large- and intermediate-$p^2$
domains.  The study used Landau gauge because it is a fixed point of the QCD
renormalisation group and $Z_2\approx 1$, even nonperturbatively\cite{mr97}.}
\begin{equation}
\label{tableA} 
\begin{array}{lccccc}
   & \bar m& b_0 & b_1 & b_2 & b_3 \\\hline
 u & 0.00948 & 0.131 & 2.94 & 0.733 & 0.185 \\
 s & 0.210   & 0.105 & 3.18 & 0.858 & 0.185   
\end{array}
\end{equation}
were fixed in a least-squares fit to light- and heavy-meson
observables~\cite{misha}, with these dimensionless $u,s$ current-quark masses
corresponding to
\begin{equation}
m_u^{1\,{\rm GeV}} = 5.4\,{\rm MeV},\;
m_s^{1\,{\rm GeV}}=119\,{\rm MeV}\,.  
\end{equation}
This algebraic parametrisation combines the
effects of confinement and dynamical chiral symmetry breaking with
free-particle behaviour at large spacelike $p^2$~\cite{echaya}.

The dominant component of the $\pi$- and $K$-meson Bethe-Salpeter amplitudes
is primarily determined by the axial-vector Ward-Takahashi
identity\cite{mr97,mrt98}: 
\begin{equation} 
\label{piKamp}
\Gamma_H(k^2)  =  i\gamma_5\,\case{\sqrt{2}}{f_{H}}\,B_{H}(k^2)\,,\;H=\pi,K\,,
\end{equation}
where $B_H:=\left.B_u\right|^{\bar m_u\to 0}_{b_0^u\to b_0^P}$ and
\cite{misha}
\begin{equation}
b_0^\pi = 0.204\,,\;
b_0^K = 0.319\,;
\end{equation}
i.e., $B_H$ is the quark quark mass function obtained from
Eqs.~(\ref{qprop})-(\ref{svm}) with $\bar m_f =0$ and $b_0^f$ replaced by the
values indicated.  With these dressed-propagators and Bethe-Salpeter
amplitudes one obtains (in GeV)
\begin{equation}
\label{oldresults}
\begin{array}{l|cccc}
        & f_\pi & m_\pi & f_K & m_K \\\hline 
{\rm Calc.} & 0.146 & 0.130 & 0.178 & 0.449 \\
{\rm Obs.}\cite{pdg98} & 0.131 & 0.138 & 0.160 & 0.496 
\end{array}
\end{equation}
and $\langle \bar q q \rangle^{1\,{\rm GeV}} = (0.220\,{\rm GeV})^3$.

\subsection{Neutral kaon Decay}
We now turn to the transitions $K^0_S \to \pi^+\pi^-$, $\pi^0\pi^0$.  In
comparison with $K^+\to\pi^+\pi^0$ there is a significant qualitative
difference: all effective operators contribute to these transitions and
furthermore the QCD penguin operators: $Q_{5,6}$, and ew penguin operators:
$Q_{7,8}$, can direct the transition through $0^{++}$ intermediate states.
This can be important because; e.g., Ref.~\cite{expsigma} reports evidence of
a broad scalar resonance in $\tau \to \nu_\tau \pi^- \pi^0 \pi^0$ decays:
\begin{equation}
\label{propsigma}
m_{0^{++}} \approx 1.12\,m_K\,,\;
\Gamma_{0^{++}\to\,\pi\pi}\approx 0.54\,{\rm GeV}\,,
\end{equation}
and with $m_{0^{++}} \approx m_K$ such a resonance could provide a
significant contribution to the nonleptonic $K^0$ decays.  We explore this
hypothesis by allowing such a contribution in our analysis: $K^0_S \to
\sigma_{0^{++}} \to \pi\pi$.

Before proceeding further we note that a light scalar meson is a feature of
DSE studies using a well-constrained rainbow-ladder
truncation\cite{jm93,bsesep,pieter}.  However, in the $0^{++}$ channel this
truncation is very likely unreliable\cite{cdrqcII}, a fact we expect is
entangled with the phenomenological difficulties encountered in understanding
the composition of scalar resonances below $1.4\,$GeV\cite{mikescalars}.
While the lack of a reliable DSE truncation for scalar mesons prevents an
accurate calculation of their Bethe-Salpeter amplitude and mass, they are
nevertheless describable by such amplitudes, which herein we parametrise as
\begin{equation}
\label{gsigma}
\Gamma_\sigma(k;p) = I_D\,\case{1}{{\cal N}_\sigma}\,
\frac{1}{1+ (k^2/\omega_\sigma^2)^2}\,,
\end{equation}
where $I_D= \gamma_4^2$ and $\omega_\sigma$ is a width parameter.
$\Gamma_\sigma$ is normalised canonically and consistent with the impulse
approximation ($q_\pm= q \pm \case{1}{2} p$)
\begin{eqnarray}
\nonumber   
\lefteqn{ p_\mu  = 
N_c \,{\rm tr} \int_q^\Lambda
\left[ \Gamma_\sigma(q;-p) 
\frac{\partial S(q_+)}{\!\!\!\!\!\!\partial p_\mu} 
\Gamma_\sigma(q; p) S(q_-) \right. }\\
&+ & \left.\left. 
 \Gamma_\sigma(q;-p) S(q_+) \Gamma_\sigma(q; p) 
        \frac{\partial S(q_-)}{\!\!\!\!\!\!\partial p_\mu}\right]
        \right|_{p^2=-m_\sigma^2}\! .
\label{Nsigma}
\end{eqnarray}

We separate the $Q_6$ contribution to the $K^0_S \to \pi\pi$ transition into
two parts and consider first the new class of contributions, which introduce
the $\sigma$ intermediate state:
\begin{eqnarray}
\lefteqn{\langle \pi(p_1) \pi(p_2)| Q_6 | K^0(p)\rangle = }\\
&&\nonumber \langle \pi(p_1)\pi(p_2) | \sigma(p)\rangle\,
D_\sigma(p^2)\,\langle \sigma(p) | Q_6 | K^0(p) \rangle\,,
\end{eqnarray}
where we represent $\sigma$ propagation by
\begin{equation}
\label{Dsigma}
D_\sigma(p^2) = 1/[p^2 + m_\sigma^2]
\end{equation}
and employ the impulse approximation for the $\sigma\pi\pi$
coupling
\begin{eqnarray}
\label{sigpipi}
\lefteqn{M_{\sigma\pi\pi}(p_1,p_2)  := 
\langle \pi(p_1)\pi(p_2) | \sigma(p)\rangle = }\\
&&  \nonumber
2 N_c {\rm tr}\int_k^\Lambda
\Gamma_\sigma(k;p)\,S_u(k_{++}) \\
&& \nonumber \times 
i\Gamma_\pi(k_{0+};-p_1)\,
S_u(k_{+-})\,i\Gamma_\pi(k_{-0};-p_2)\,S_u(k_{--})\,,
\end{eqnarray}
$k_{\alpha\beta}= k+(\alpha/2)p_1+(\beta/2)p_2$, which provides the basis for
the calculation of $\Gamma_{0^{++}\to\pi\pi}$.  This combination of
simple-pole propagator plus impulse approximation coupling to the dominant
decay channel is phenomenologically efficacious; e.g., Refs.~\cite{mike2}.

In impulse approximation 
\begin{eqnarray}
\lefteqn{\langle \sigma(p) | Q_6 | K^0(p) \rangle = 
\sqrt{2}\,N_c^2\,\times }\\
&& \nonumber
{\rm tr}\,Z_4^2
\int_{k_1}^\Lambda\!\int_{k_2}^\Lambda
i\chi_K(k_1;\case{1}{2}p,\case{1}{2}p)\,
O_\mu^+\,\chi_\sigma(k_2;-\case{1}{2}p,-\case{1}{2}p)\,O_\mu^-\,,
\end{eqnarray}
with $Z_4(\mu,\Lambda)$ the mass renormalisation constant and
$\chi_\sigma(k;\ell_1,\ell_2)$ an obvious analogue of
$\chi_\pi(k;\ell_1,\ell_2)$ in Eq.~(\ref{chipi}).  Using
\begin{equation}
{\rm tr}[G_1 O_\mu^+ G_2 O_\mu^-] = 2\, {\rm tr}[G_1 (1-\gamma_5)]\, {\rm
tr}[G_2 (1+\gamma_5)] 
\end{equation}
this yields
\begin{eqnarray}
\nonumber 
\lefteqn{-\case{1}{\sqrt{2}}\langle \sigma(p) | Q_6 | K^0(p) \rangle = }\\
&& \nonumber
 \left(\sqrt{2} \,N_c\,
{\rm tr}Z_4\!\int_{k_1}^\Lambda i\gamma_5\,
\chi_K(k_1;\case{1}{2}p,\case{1}{2}p)\right) \\
&\times& 
\left(\sqrt{2}\,N_c\,{\rm tr}Z_4\!\int_{k_2}^\Lambda 
\chi_\sigma(k_2;-\case{1}{2}p,-\case{1}{2}p)\right)\,.
\end{eqnarray}
From Refs.~\cite{mr97,mrt98} we identify the first parenthesised term as the
residue of the kaon pole in the pseudoscalar vertex:
\begin{eqnarray}
\nonumber \lefteqn{i r_K:= }\\
&& \sqrt{2}\,N_c\,{\rm tr}Z_4\!\int_{k_1}^\Lambda\gamma_5\,\
\chi_K(k_1;\case{1}{2}p,\case{1}{2}p)
= \frac{f_K\,m_K^2}{m_u + m_s}\,.
\end{eqnarray}
The second term is the scalar meson analogue in the scalar vertex but the
vector Ward-Takahashi identity, which is relevant in this case, does not make
possible an algebraic simplification.  The integral and its $\mu$-dependence
must therefore be calculated.  That is straightforward when the
renormalisation-group-improved rainbow-ladder truncation is accurate; e.g.,
Refs.~\cite{mt99,mr97,jm93}, but not yet for scalar mesons.  This is where
the simple {\it Ansatz} of Eq.~(\ref{gsigma}) is useful: it yields a finite
integral and we therefore suppress $Z_4$ to obtain
\begin{eqnarray}
\lefteqn{\case{1}{\sqrt{2}}\langle \sigma(p) | Q_6 | K^0(p) \rangle }\\
& = & \nonumber 
r_K \,
\sqrt{2}\,N_c\,{\rm tr}\int_{k_2}^\Lambda
\chi_\sigma(k_2;-\case{1}{2}p,-\case{1}{2}p) 
=: r_K\, r_\sigma(p^2)\,.
\end{eqnarray}
The result for $Q_5$ is similar, but suppressed by a factor of $1/N_c$, and
the contribution of the ew penguins: $Q_{7,8}$, can be obtained similarly.

The other class of contributions, which do not involve a $0^{++}$
intermediate state, can be evaluated following the explicit example of $Q_1$
presented above.  Only two additional three-point functions arise:
\begin{eqnarray}
{\cal G}_\pi^S(p_1,p_2) & = & 
\langle \pi(p_1)\pi(p_2)|(\bar u u + \bar d d)|0\rangle\,,\\
-{\cal G}_{K\pi}^S(p_1,p_2) & = & 
\langle \pi^-(p_1)|\,\bar s u\, |K^0(p)\rangle \,.
\end{eqnarray}
They are the scalar pion form factor and the scalar $K\pi$ transition form
factor, respectively, and ${\cal G}_{K\pi}^S(p_1,p_2)$ can be expressed
without additional calculation in terms of the $K_{\ell 3}$ form
factors\cite{kl3}:
\begin{eqnarray}
\lefteqn{{\cal G}_{K\pi}^S(p_1,p_2) = }\\
&& \nonumber
\frac{p_1^2 - p_2^2}{m_s - m_d}\,
\left[ f_+^K(-p_2^2) + \frac{p_2^2}{p_2^2-p_1^2} \,f_-^K(-p_2^2)
\right]\,,
\end{eqnarray}
a result which follows from the vector Ward-Takahashi identity.  A
preliminary result is available\cite{isaac} for ${\cal G}_\pi^S(p_1,p_2)$,
which takes the form anticipated from current algebra.  That is to be
expected because correctly truncated DSE models provide a good description of
chiral symmetry and its dynamical breakdown, as illustrated in a study of
$\pi\pi$ scattering\cite{pipi}.  This makes a calculation of ${\cal
G}_\pi^S(p_1,p_2)$ unnecessary for our present study because we can adopt the
form\cite{gl84} [$(r_\pi^S) = 3.76\,$GeV$^{-1}$]:
\begin{equation}
{\cal G}_\pi^S(p_1,p_2) = - 4 \frac{\langle \bar q q \rangle}{f_\pi^2}\,
\left[ 1 - \case{1}{6} \, (r_\pi^S)^2 \,(p_1+p_2)^2 \right]\,.
\end{equation}
 
The matrix elements for the $K\to \pi\pi$ transitions can all be written
\begin{eqnarray}
\label{genform}
M_{K\to\pi\pi} & = & M_{K\to\pi\pi}^{\rm qcd} + \alpha_{\rm em}
\,M_{K\to\pi\pi}^{\rm ew}\,,
\end{eqnarray}
with the explicit forms given in the appendix and the pure isospin amplitudes
defined in Eqs.~(\ref{A0}), (\ref{A2}).

\section{Results and Discussion}
\label{sec3}
Everything required for our calculation of the widths is now specified.
There are two parameters: $\omega_\sigma$ in Eq.~(\ref{gsigma}); and
$m_\sigma$ in Eq~(\ref{Dsigma}).  We determine them in a least-squares fit
to: $\Gamma_{K^0_S\to \pi^+\pi^-}$, $\Gamma_{K^0_S\to \pi^0\pi^0}$, taken
from Ref.~\cite{pdg98}; and $\Gamma_{\sigma \to (\pi\pi)}$ in
Eq.~(\ref{propsigma}), and obtain
(in GeV)
\begin{equation}
\label{results}
\begin{array}{l|cc}
                                & {\rm Obs.}    & {\rm Calc.} \\\hline
m_\sigma                        & 1.12\,m_K     & ~~~~1.14\,m_K\\
\omega_\sigma                   &               & 0.611   \\\hline
\Gamma_{\sigma\to (\pi\pi)}     &  0.54         & 0.54 \\
\Gamma_{K^0_S\to \pi^+\pi^-}^{\times 10^{-15}}
                                & 5.055\pm 0.025& 5.16 \\
\Gamma_{K^0_S\to \pi^0\pi^0}^{\times 10^{-15}}
                                & 2.305\pm 0.023& 2.11 \\
\Gamma_{K^+\to \pi^+\pi^0}^{\times 10^{-15}}
                                & 0.0112\pm 0.0001 & ~~~0.0116 
\end{array} 
\end{equation}
which is a relative error on fitted quantities of $< 4$\%.\footnote{We used
$G_F= 1.166\times 10^{-5}\,$GeV$^{-2}$, $V_{ts}= 0.0385$, $V_{td}=0.0085$,
$V_{us}= 0.220$, $V_{ud}=0.975$, Im$(V_{ts}^\ast V_{td})=0.000133$, and $c_i$
obtained from Eq.~(\protect\ref{coeffs}).  With $\Gamma_\sigma(k;p)\propto
\exp(-k^2/\omega_\sigma^2)$ instead of Eq.~(\ref{gsigma}), $m_\sigma =
1.12\,m_K$, $\omega_\sigma=0.694\,$GeV yields exactly the same results for
the calculated quantities.}
This value of $\omega_\sigma$ corresponds to an {\it intrinsic}
$\sigma$-meson size: $r^I_\sigma:= 1/\omega_\sigma$, which is $0.84\,
r^I_\rho$; i.e, $84$\% of that of the $\rho$-meson determined in
Ref.~\cite{misha}.  

The widths in Eq.~(\ref{results}) are obtained from the calculated amplitudes
(in GeV with $m_K$ from Eq.~(\ref{oldresults}))
\begin{eqnarray}
|M_{K^0\to\pi^+\pi^-}| & = & 2.7 \times 10^{-7} = 5.9 \times 10^{-7}\,m_K\,, \\
|M_{K^0\to\pi^0\pi^0}| & = & 2.4 \times 10^{-7} = 5.4 \times 10^{-7}\,m_K\,, \\
|M_{K^+\to\pi^+\pi^0}| & = & 1.8 \times 10^{-8} = 4.0 \times 10^{-8}\,m_K \,.
\end{eqnarray}
For the pure isospin amplitudes we find (in GeV):
\begin{equation}
{\rm Re}(A_0)  =  31.7 \times 10^{-8}\,,\;
{\rm Re}(A_2)  =  1.47 \times 10^{-8} \,,
\end{equation}
which are consistent with recent lattice estimates\cite{crzylattice} and
yield
\begin{equation}
1/w = 21.6\,.
\end{equation}

Our analysis also yields values of the parameters: $B_i^{(1/2),(3/2)}\!$,
used in phenomenological analyses to express the operator expectation
values\cite{buras}.  Of course, $B_1^{(3/2)} = 0.5$, as discussed in
connection with Eq.~(\ref{b1fix}) and, using the formulae in the appendix, we
obtain algebraically: $B_1^{(1/2)} = B_2^{(1/2)} = B_3^{(1/2)} = B_4^{(1/2)}
= B_2^{(3/2)}= B_1^{(3/2)}$.  We also calculate
\begin{eqnarray}
B_5^{(1/2)} = B_6^{(1/2)} & = & 1.43 + (17.9)_\sigma\,,
\end{eqnarray}
where the second term is the contribution of the $\sigma$-meson.  The
non-$\sigma$ contribution is large because of the strength of the $K\pi$
transition form factor.  If the vacuum saturation {\it Ansatz} is used to
estimate the operator expectation values they are all $\equiv 1$.  That
method does not admit a $\sigma$-meson contribution.

Eliminating the ew penguin contributions yields a $<1$\% reduction in $1/w$,
which is consistent with the the magnitude of $\alpha_{\rm em}$.  Suppressing
instead the $\sigma$-meson contribution, while not affecting
$\Gamma_{K^+\to\pi^+\pi^0}$ of course (see Eq.~(\ref{a1})), yields
$\Gamma_{K^0_S\to \pi^+\pi^-}= 1.3\times 10^{-16}\,$GeV, $\Gamma_{K^0_S\to
\pi^0\pi^0} = 1.1 \times 10^{-17}\,$GeV, and $1/w=2.9$.

The value of $\epsilon^\prime/\epsilon$ follows from Eq.~(\ref{ratio}).
Suppressing the $\sigma$-meson and ew penguin contributions we obtain
$\epsilon^\prime/\epsilon = 128\times 10^{-3}$, which is $\sim 60$-times
larger than the experimental average in Eq.~(\ref{expt}).  Including the
$\sigma$-meson we find $31.3\times 10^{-3}$.  To understand these results we
note that Eq.~(\ref{ratio}) can be written
\begin{eqnarray}
\label{newgen}
\frac{\epsilon^\prime}{\epsilon} & = &
-\case{1}{\sqrt{2}}\frac{w}{|\epsilon|}
\frac{{\rm Im} A_0}{{\rm Re} A_0}
\left\{ 1 - \frac{1}{w} \frac{{\rm Im} A_2}{{\rm Im}A_0}\right\}\,,
\end{eqnarray}
which makes clear that the ratio is determined by Im$(A_0)$/Re$(A_0)$ {\it
unless} Im$(A_2)\neq 0$.  Noting that $c_{1,2}$ are real, Eq.~(\ref{coeffs}),
then it follows from Eq.~(\ref{a1}) that Im($A_2)=0$ in the absence of ew
effects.  Hence our calculated results are large because the pre-factor in
Eq.~(\ref{newgen}) is large.  The dependence on the $\sigma$ contribution is
easily understood.  The pre-factor is $\propto {\rm Im}(A_0)/{\rm
Re}(A_0)^2$, which is large in the absence of the $\sigma$ contribution even
though Im$(A_0)$ and Re$(A_0)$ are individually small.  The $\sigma$
contribution adds simultaneously to Im($A_0$) and Re$(A_0)$ with a magnitude
$\sim 100$-times larger than the original values.  Hence the final ratio is
sensitive only to the relative strength of the $\sigma$ contributions, which
is determined by the coefficients $c_{5,6}$.

Including both the $\sigma$ and ew penguin contributions we obtain
\begin{equation}
\epsilon^\prime/\epsilon = 31.7 \times 10^{-3}\,,
\end{equation}
from which it is clear that the ew penguins are a correction of order
$\alpha_{em}$ as one would naively expect.  In this case Im($A_2)\neq 0$.
However, as observed above, the $\sigma$-meson enhancement responsible for
the $\Delta I=\case{1}{2}$ rule affects the real and imaginary parts of $A_0$
simultaneously so that $(1/w)\,{\rm Im} A_2/{\rm Im}A_0$ remains negligible.

If we employ the artifice of an {\it ad hoc} suppression of the $\sigma$
contribution to Im($A_0)$ while retaining it in Re$(A_0)$; i.e., 
make the replacement
\begin{eqnarray}
c_i\, {\cal M}_3 & \to & {\rm Re}(c_i\,{\cal M}_3),\;
i=5,6,7,8
\end{eqnarray}
in Eqs.~(\ref{a1})-(\ref{a6}), we find
\begin{equation}
\label{ratioadhoc}
\epsilon^\prime/\epsilon  = 2.7 \times 10^{-3}\,.
\end{equation}
This artifice is implicit in the phenomenological analyses reviewed in
Ref.~\cite{buras} and that is why Eq.~(\ref{ratioadhoc}) reproduces their
order-of-magnitude.  The small value is only possible because in this case
Im$(A_0)$ is not $\sigma$-enhanced and is therefore of the same magnitude as
${\rm Im}A_2\,/w \propto \alpha_{\rm em}/w$, due to the $1/w$ enhancement
factor.  That factor survives because \underline{Re($A_0)$} is still
magnified as required in order to satisfy the $\Delta I=\case{1}{2}$ rule.
Currently we cannot justify this procedure.\footnote{NB. If this procedure is
followed then $m_u\neq m_d$ isospin symmetry breaking effects also contribute
significantly to $\epsilon^\prime/\epsilon$.}

\section{Epilogue}
\label{epil}
We have demonstrated that estimating the $K\to \pi\pi_{I=2}$ matrix element
using the impulse approximation is algebraically equivalent to using the
vacuum saturation {\it Ansatz} and yields a result that is $\sim 2$-times too
large.  The identification of a compensating mechanism that can correct for
this overestimate is a contemporary challenge.  

We have also shown that the contribution of a light scalar meson mediated by
the QCD penguin operators: $Q_{5,6}$, is a plausible candidate for the
long-range mechanism underlying the enhancement of $K\to \pi\pi_{I=0}$
transitions.\footnote{$Q_{5,6}$ mediated scalar diquark transitions:
$(us)_{0^+}^{I=1/2}\to (ud)_{0^+}^{I=0}$, are the $s\to t$-channel
interchange of the interaction that herein produces the $\sigma$-meson.  They
are a viable candidate for the mechanism that produces the $\Delta
I=\case{1}{2}$ enhancement for baryons.  This was explored in
Ref.~\protect\cite{stech}, however, the requirement therein that diquarks
also explain the enhancement for mesons appears unnecessarily cumbersome.}
A good description of that enhancement {\it requires} a mass and width for
this $0^{++}$ resonance that agree with those recently
inferred\cite{expsigma}, and the analysis is not sensitive to details of the
model Bethe-Salpeter amplitude.  However, this same mechanism yields a value
of $\epsilon^\prime/\epsilon$ that is $\sim 15$-times larger than the average
of contemporary experimental results unless a means is found to suppress its
contribution to Im($A_0)$.

If a light scalar resonance exists it will contribute in the manner we have
elucidated and should be incorporated in any treatment of $K\to \pi\pi$.

\section*{Acknowledgments}
M.A.I.  acknowledges the hospitality and support of the Department of Physics
at VPI and the Physics Division at ANL during visits in which this work was
conducted.  This work was supported by the US Department of Energy, Nuclear
Physics Division, under contract numbers: DE-FG02-96ER40994 and
W-31-109-ENG-38, and benefited from the resources of the National Energy
Research Scientific Computing Center.  S.M.S. is grateful for financial
support from the A.v. Humboldt foundation.

\appendix
\section*{Collected Formulae}
The matrix elements for the $K\to \pi\pi$ transitions are all of the form in
Eq.~(\ref{genform}) with
\begin{eqnarray}
\label{a1}
M_{K^+\to\pi^+\pi^0}^{\rm qcd} & = & 
\case{1}{\sqrt{2}}\,\tilde G_F\,(1+\case{1}{N_c})\,(c_1 + c_2)\,{\cal M}_1\,,\\
\nonumber
M_{K_S^0\to\pi^+\pi^-}^{\rm qcd} & = & 
\tilde G_F\,\left\{[
c_2 + c_4 + \case{1}{N_c}(c_1+c_3)]{\cal M}_1 
\right.\\
& & \left.
+ 2 \,( \case{1}{N_c} c_5 + c_6)\, ({\cal M}_2 + \case{1}{\sqrt{2}} {\cal M}_3)
\right\}\,,\\
\nonumber
M_{K_S^0\to\pi^0\pi^0}^{\rm qcd} & = & 
\tilde G_F\,\left\{[
c_4 - c_1 - \case{1}{N_c}(c_2-c_3)]{\cal M}_1 
\right.\\
& & \left.
+ 2 \,( \case{1}{N_c} c_5 + c_6)\, ({\cal M}_2 + \case{1}{\sqrt{2}} {\cal M}_3)
\right\}\,,
\end{eqnarray}
\begin{eqnarray}
\label{a4}
\lefteqn{M_{K^+\to\pi^+\pi^0}^{\rm ew}  = 
 - \case{1}{\sqrt{2}}\tilde G_F\,\left\{
\case{3}{2}\left[c_7 + \case{1}{N_c} c_8 \right.
\right.
}\\
&& \nonumber 
\left.
\left.
- (1+\case{1}{N_c}) (c_9+c_{10}) \right]{\cal M}_1
+ 3 (\case{1}{N_c} c_7 + c_8)\,{\cal M}_2^b\right\}\,,\\
\lefteqn{M_{K_S^0\to\pi^+\pi^-}^{\rm ew}  = \tilde G_F\,
\left\{ (\case{1}{N_c} c_9 + c_{10}) {\cal M}_1  \right.
}\\
&& \nonumber
\left.
- (\case{1}{N_c} c_7 + c_8) \, ({\cal M}_2^a+ 2 {\cal M}_2^b 
+ \case{1}{\sqrt{2}} {\cal M}_3 )
\right\}\,,\\
\label{a6}
\lefteqn{M_{K_S^0\to\pi^0\pi^0}^{\rm ew}  = \tilde G_F\,
\left\{ \left[ c_7 + \case{1}{N_c} c_8 - (1 + \case{1}{2 N_c}) c_9 
\right. \right. }\\
&& \nonumber
\left. \left. 
- \case{1}{2} (1 + \case{2}{N_c}) c_{10} \right] {\cal M}_1
- ( \case{1}{N_c} c_7 + c_8 ) ({\cal M}_2 
+ \case{1}{\sqrt{2}} {\cal M}_3)
\right\}\,,\\
{\cal M}_2^a& = & r_K\,{\cal G}_\pi^S(p_1,p_2)\,,\;
{\cal M}_2^b  =  r_\pi\,{\cal G}_{K\pi}^S(p_1,p_2)\,,\\
{\cal M}_2 & = & {\cal M}_2^a- {\cal M}_2^b\,,\\ 
{\cal M}_3 & = & r_K\,r_\sigma(p^2)\,D_\sigma(p^2)\,
M_{\sigma\pi\pi}(p_1,p_2)\,,
\end{eqnarray}
with $p^2=(p_1+p_2)^2=-m_K^2$, $p_1^2=p_2^2=-m_\pi^2$.  These formulae make
clear the operators that would be suppressed if $N_c$ were large.  Note that
\begin{equation}
M_{K^+\to\pi^+\pi^0}^{\rm qcd} = \case{1}{\sqrt{2}}
(M_{K_S^0\to\pi^+\pi^-}^{\rm qcd}  - M_{K_S^0\to\pi^0\pi^0}^{\rm qcd} )\,.
\end{equation}
This is not true of the complete amplitude.

In our calculations we use values of the coefficients that correspond to our
choice of $\Lambda_{\rm QCD}\sim 0.2\,$GeV: $c_i= z_i + \tau\,y_i$, $\tau= -
(V^\ast_{ts} V_{td})/(V_{us}^\ast\,V_{ud})$ with\cite{buras}
\begin{equation}
\label{coeffs}
\begin{array}{c|rr}
      &  z_i~~   &  y_i~~~ \\\hline
1     & -0.407 & 0.0~~~ \\
2     &  1.204 & 0.0~~~ \\
3     &  0.007 & ~0.023 \\
4     & -0.022 &-0.046 \\
5     &  0.006 & ~0.004 \\
6     & -0.022 &-0.076 \\
7     &  0.003 &-0.033 \\
8     &  0.008 & ~0.121 \\
9     &  0.007 &-1.479 \\
10    & -0.005 & ~0.540
\end{array}
\end{equation}
Using the alternative set listed in Ref.~\cite{buras} then, with $m_\sigma=
1.06\,m_K$ and $\omega_\sigma= 0.670\,$GeV, we obtain results that differ
from those in Eq.~(\ref{results}) by $\lsim 1$\%, and
$\epsilon^\prime/\epsilon = 69.0\times 10^{-3}$ primarily because $y_5$ in the
alternative set is $2.6$-times as large.

From the complete matrix elements: Eqs.~(\ref{genform}) and
Eqs.~(\ref{a1})-(\ref{a6}), we obtain the widths:
\begin{eqnarray}
\label{kpppp0}
\Gamma_{K^+\to\pi^+\pi^0} & = & 
        {\cal C}(m_K)\,|M_{K^+\to\pi^+\pi^0}|^2\,,\\
\Gamma_{K^0_S\to\pi^+\pi^-} & = & 
       2\, {\cal C}(m_K)\,|M_{K_S^0\to\pi^+\pi^-}|^2\,,\\
\Gamma_{K^0_S\to\pi^0\pi^0} & = & 
        {\cal C}(m_K)\,|M_{K_S^0\to\pi^0\pi^0}|^2\,,\\
{\cal C}(x) & = & \frac{1}{16\pi\,x} \sqrt{1 - \case{4m_\pi^2}{x^2}}\,,
\end{eqnarray}
while the matrix element of Eq.~(\ref{sigpipi}) features in
\begin{eqnarray}
\Gamma_{\sigma\to(\pi\pi)} & = & \case{3}{2}\,{\cal C}(m_\sigma)\,
        |M_{\sigma\pi\pi}(m_\sigma^2;m_\pi^2,m_\pi^2)|^2\,.
\end{eqnarray}


%
\end{document}